\documentclass[aps,prl,reprint,superscriptaddress,nofootinbib]{revtex4-2}
\usepackage{graphicx}
\usepackage{amsmath,amssymb}
\usepackage{xcolor}
\usepackage[colorlinks=true,allcolors=blue]{hyperref}

\newcommand{\GeV}{\,\mathrm{GeV}}
\newcommand{\TeV}{\,\mathrm{TeV}}
\newcommand{\mm}{\,\mathrm{mm}}
\newcommand{\ps}{\,\mathrm{ps}}
\newcommand{\pb}{\,\mathrm{pb}}
\newcommand{\fbinv}{\,\mathrm{fb}^{-1}}

\newcommand{\ctau}{c\tau}
\newcommand{\ms}{m_s}
\newcommand{\dTSV}{\Delta T_{\rm SV}}
\newcommand{\RSV}{R_{\rm SV}}
\newcommand{\flate}{f_{\rm late}}
\newcommand{\sigmat}{\sigma_t}
\newcommand{\Gain}{\mathcal{G}}

\begin{document}

\title{
Vector-boson-fusion timing references for four-dimensional displaced-vertex searches
}

\author{Renjie Wang}
\email{rjwang@ihep.ac.cn}
\affiliation{Institute of High Energy Physics,
  Chinese Academy of Sciences, Beijing 100049, China}

\date{\today}

\begin{abstract}
Long-lived particles that decay inside the tracker are searched for via the spatial displacement of their decay vertices. However, the same decays also arrive late, providing a handle made available by a precision timing layer at the HL-LHC that displaced-vertex searches do not exploit. We study the exotic decay of a Higgs boson produced through vector-boson fusion (VBF), $pp\to hjj$ with $h\to ss$, in which the long-lived scalar decays as $s\to b\bar b$, and use the VBF tag jets to identify the hard-scatter vertex. Its prompt central tracks, timed by a CMS-MTD-like barrel layer that also times the central tracks of the displaced vertex, define a per-event start time, turning each displaced vertex into a four-dimensional object. Because a fast simulation cannot determine the absolute normalization of the heavy-flavour and instrumental displaced-vertex backgrounds, we take as the primary result a normalization-free figure of merit: the ratio of the timing-enhanced sensitivity to the spatial-only sensitivity. Adding the timing layer improves this ratio by a factor of $\simeq2.8$ at a nominal working point and by up to an order of magnitude for a tighter delayed-vertex requirement, because the heavy-flavour background that survives the spatial selection is prompt in collision time. The improvement increases with decay length, extending sensitivity into the long-lifetime regime in which purely spatial searches lose tracker acceptance. The gain is essentially mass-independent at a decay length of $\ctau\simeq100\mm$, where the trend with scalar mass reverses: lighter scalars benefit more at short lifetimes because of their larger boost, whereas heavier scalars benefit more at long lifetimes as the lighter states leak out of the tracker.
\end{abstract}

\maketitle

\textit{Introduction.}---Exotic Higgs decays to long-lived scalars,
$h\to ss$ with $s\to b\bar b$, are among the most well-motivated signatures of a
hidden sector coupled to the SM through the Higgs
portal~\cite{Strassler:2006im,Strassler:2006ri,Craig:2015pha,Curtin:2013fra,Curtin:2015fna,Bhattacherjee:2021zvv,Cepeda:2019klc}.
Dedicated searches at the LHC reconstruct the displaced secondary vertex (SV)
produced by the scalar decay from its spatial displacement
alone~\cite{Alimena:2019zor,Lee:2018pag}.
Recent ATLAS and CMS analyses have pushed these purely spatial searches to high
sensitivity: the ATLAS Run-2 displaced-vertex search~\cite{ATLAS:2024grv} targets
hadronic long-lived particles (LLPs) in the mass range $\ms\simeq5\text{--}55\GeV$,
includes a VBF-tagged signal region, and sets the strongest current limits for
$\ctau\lesssim100\mm$; the CMS Run-3 light-LLP displaced-jets
search~\cite{CMS:2024lowmass} achieves up to an order-of-magnitude improvement
over earlier results for $\ms\lesssim60\GeV$ and $\ctau\lesssim1\,\mathrm{m}$ through
dedicated triggers, improved reconstruction, and machine-learning
discriminants~\cite{CMS:2021juv,ATLAS:2021jig}. In their most sensitive
displacement-limited regions, however, these analyses are already constrained by
the systematic uncertainty of the data-driven background estimate rather than by
the available luminosity: additional spatial information or luminosity alone cannot
break this ceiling.
Displacement provides only half of the kinematic information carried by a long-lived decay:
a particle that travels a macroscopic distance before decaying also
produces tracks that arrive late, and this delay grows with the same
flight distance that the spatial search exploits.

Precision timing turns that delay into an independent observable. The arrival
time of long-lived decay products has been proposed as a means to reconstruct LLP kinematics
and lifetimes~\cite{Flowers:2019eoc,Banerjee:2019ktv,Cerri:2018rkm}, to trigger
on displaced jets~\cite{Bhattacherjee:2021rml}, and, in the pioneering study of
Ref.~\cite{Liu:2018wte}, to suppress the prompt background to a neutral
long-lived particle by an order of magnitude using a $\mathcal{O}(30\ps)$ timing
layer at the HL-LHC~\cite{Cerri:2018skj}. That study used an
initial-state-radiation jet as the timing reference, leaving open the question of how to obtain a
clean, reproducible start time $t_0$ in the realistic high-pileup environment of
the HL-LHC, where an ISR jet is neither guaranteed nor unambiguously prompt.
The present work is not the first to propose timing for long-lived particles; its
novelty lies in identifying a specific configuration in which the event start time
is provided by the same objects that define the signal region, making it
available event by event and tying it unambiguously to the hard scatter.

Vector-boson fusion provides exactly this configuration. The two forward,
high-mass tag jets that strengthen the spatial selection also tag the hard-scatter
vertex with high purity; VBF has previously been used as a production and trigger
handle for long-lived particles~\cite{Jones-Perez:2019plk}, but here it plays a
new role as the enabler of the timing reference. The CMS Phase-2 upgrade adds a MIP
Timing Detector (MTD)~\cite{CMS-TDR-020}, whose Barrel Timing Layer (BTL,
$|\eta|<1.48$) timestamps charged tracks with a resolution of
$\simeq30\text{--}60\ps$; its detailed parameters are given below. Such a
CMS-MTD-like barrel layer times the prompt central tracks of the VBF-tagged
vertex to determine $t_0$, and the same layer times the central tracks of the displaced
vertex under study. The VBF requirement that defines the signal region is
therefore identical to the requirement that provides the temporal reference. The
VBF jets themselves supply no timing measurement; the reference is instead the production
time of the prompt central tracks of the vertex they tag. A realistic HL-LHC
analysis would combine several complementary handles---prompt-track timing,
primary-vertex time, calorimeter or jet timing, and object-to-vertex
association~\cite{Gligorov:2026review}---of which we deliberately isolate the
single cleanest one here.

In this Letter we present, to our knowledge, the first study to use the
VBF-tagged hard-scatter vertex as the event-by-event timing reference for
four-dimensional (4D) displaced-vertex reconstruction of long-lived scalars at
the HL-LHC. Our aim is not to reproduce the full experimental searches, whose
background estimates are necessarily data driven, but to isolate the incremental
information carried by vertex time once a spatial displaced-vertex candidate has
been reconstructed. Because a fast detector simulation cannot determine the absolute
background normalization, we quote a normalization-free figure of merit---the
gain in $S/\sqrt{B}$ from adding timing to an otherwise identical spatial
selection---and map it across the scalar mass--lifetime plane. To place this
gain in physical context, we also translate the selection into illustrative,
statistics-limited 95\% confidence-level (CL) limits on the branching ratio
$\mathrm{BR}(h\to ss)$---under the explicit
assumption $\mathrm{BR}(s\to b\bar b)=1$, equivalently a limit on the product
$\mathrm{BR}(h\to ss)\,[\mathrm{BR}(s\to b\bar b)]^2$ for this topology---using
the standard asymptotic modified-frequentist ($\mathrm{CL}_s$) procedure. The
central result is a
structural one: the timing gain reverses its dependence on scalar mass at a decay
length of $\ctau\simeq100\mm$, the point at which boost-driven and
containment-driven effects balance, and it grows with lifetime precisely where
purely spatial searches lose tracker acceptance.

\textit{Signal topology and simulation.}---We study VBF Higgs production at the
$14\TeV$ HL-LHC, $pp\to hjj$, followed by the prompt exotic decay $h\to ss$ and
the displaced decay $s\to b\bar b$ of the long-lived scalar $s$, which travels a
macroscopic distance before decaying. Each event contains two
forward, high-mass tag jets recoiling against a Higgs boson whose decay yields
one or two displaced $b\bar b$ vertices. The tag jets originate from the
hard-scatter vertex and identify it with high purity; the prompt central tracks of
that vertex, timed by the barrel layer~\cite{CMS-TDR-020}, determine $t_0$, while the
same barrel layer times
the central tracks of the displaced vertex. The more forward coverage of an
endcap timing detector~\cite{CERN-LHCC-2020-007} is not required, because the
displaced $b\bar b$ tracks recoiling against the forward tag jets are themselves
central.

Events are generated with \textsc{MadGraph5\_aMC@NLO}~\cite{Alwall:2014hca} using
the PDF4LHC21 set~\cite{PDF4LHCWorkingGroup:2022cjn}, showered and hadronized in
\textsc{Pythia}~8.3~\cite{Bierlich:2022pfr}, and passed through a
\textsc{Delphes}~3~\cite{deFavereau:2013fsa} HL-LHC parametrization with a
CMS-MTD-like barrel timing layer~\cite{Addesa:2025kyl} ($|\eta|<1.48$, $\sigmat=30\ps$,
$p_T>0.7\GeV$) and pileup at $\langle\mu\rangle=200$, consistent with the Phase-2 tracker environment~\cite{CMS:Phase2TrackerTDR}. Jets are clustered with the
anti-$k_T$ algorithm~\cite{Cacciari:2008gp} ($R=0.4$) in
\textsc{FastJet}~\cite{Cacciari:2011ma}. We scan
$\ctau\in\{1,10,30,100,300,1000\}\mm$ and
$\ms\in\{20,40,55,60\}\GeV$, with the upper end approaching the $\ms\to m_h/2$
threshold. The signal and SM-Higgs samples are generated independently from the
same run card; both carry the same generated cross section after the
parton-level VBF acceptance, $\sigma_{\rm gen}=0.1725\pb$, corresponding to
$N_{\rm gen}\simeq5.2\times10^{5}$ at $\mathcal{L}=3000\fbinv$. The signal and
the electroweak and SM-Higgs backgrounds comprise $2\times10^{5}$ generated events
each; the dominant QCD background is generated with $2.2\times10^{6}$ events to
control the statistics of the late-time tail. For the gain figure of merit, all
production normalizations cancel exactly. For the illustrative branching-fraction
projection, we normalize the VBF Higgs yield to the LHC Higgs Cross Section Working
Group $14\TeV$ recommendation~\cite{deFlorian:2016spz} and use the
generator-level VBF filter only to evaluate the acceptance, while the QCD
$b\bar bjj$ continuum is treated as a nuisance parameter $\kappa_b$ rather than as
a prediction of the leading-order simulation (see below).

\textit{Event selection.}---We require at least two jets with
$p_T^{j_1}>100\GeV$ and $p_T^{j_2}>80\GeV$, invariant mass $m_{jj}>1200\GeV$,
and a rapidity gap $|\Delta\eta_{jj}|>4.0$. These cuts select the forward dijet
configuration characteristic of VBF and suppress the gluon-initiated QCD continuum, whose
central, lower-mass dijets rarely satisfy such a large gap; they accept $27.0\%$ of
generated signal at $\ms=55\GeV$ and a comparable fraction of the SM-Higgs
background, as expected for two processes sharing VBF production kinematics.
These requirements are chosen to be compatible with HL-LHC VBF-trigger
strategies~\cite{Acosta:2021qpx,Bhattacherjee:2021rml,Jones-Perez:2019plk,CMS:Phase2L1TriggerTDR,CMS:DisplacedVertexTrackTriggerPhase2L1} and are
independent of the displaced decay products, so the timing information is
used offline in this baseline study. The offline thresholds sit well above those
of a Phase-2 L1 VBF di-jet seed~\cite{CMS:Phase2L1TriggerTDR} (two jets with
$p_T\gtrsim40\GeV$ and $m_{jj}\gtrsim620\GeV$): emulating such a seed retains
$\gtrsim99.9\%$ of the offline-selected signal, placing it on the trigger
plateau, while a displaced-jet trigger~\cite{Bhattacherjee:2021rml} offers a
complementary path.

Displaced-vertex candidates are built from reconstructed tracks assigned to the
hard-scatter vertex with $p_T>1\GeV$ and transverse impact parameter
$|d_0|>0.1\mm$, clustered by the proximity of their reconstructed points of
closest approach with a single-linkage algorithm in three dimensions (linking
distance $2\mm$). Every cluster of at least three tracks whose reconstructed
transverse radius $\RSV$---the transverse distance of the vertex from the beam
line---exceeds $3\mm$ is a vertex candidate, and the highest-invariant-mass
candidate among those passing the displacement requirement is taken as the
displaced vertex of the event. Because $h\to ss$
yields two long-lived scalars, an event is retained if at least one $s$ decays
inside the tracker and forms such a vertex; a second candidate, when present, is
not used, which keeps the selection inclusive and defines the efficiency per
event rather than per scalar. Requiring two displaced vertices would suppress the
background further but would forfeit acceptance at long lifetime, where one
scalar frequently escapes the tracker. This
displacement-first choice is essential under pileup: selecting the highest-mass
cluster overall instead picks a prompt, high-multiplicity pileup combination that
then fails the displacement cut, simultaneously decimating the signal and driving
the surviving background artificially to zero. We then require $\sum p_T>10\GeV$
for the vertex tracks and a mass-to-spread ratio
$m_{\rm SV}/\Delta R_{\rm max}>4\GeV$. The $\RSV$ requirement is the single most
discriminating cut against prompt heavy flavour: it removes the majority of the
QCD vertices, whose $b$- and $c$-hadron decay lengths are sub-millimetre, while
retaining the genuinely displaced signal. Heavy flavour is exploited implicitly
here, through the vertex mass, rather than through prompt $b$-tagging: the signal
vertex is a heavy, high-multiplicity object built from the full $s\to b\bar b$
system, whereas a QCD vertex is a single, lighter $b$-hadron decay, so
$m_{\rm SV}/\Delta R_{\rm max}$ already encodes the flavour information; standard
$b$-taggers, calibrated for prompt jets, do not apply at the macroscopic
displacements relevant here.

The only Monte Carlo truth input to the reconstruction is the \textsc{Delphes}
pileup track flag, used to assign tracks to the hard-scatter vertex~\cite{CMS:2014pgm,Bertolini:2014bba}. The
selection is essentially insensitive to this idealization (quantified as the
\emph{Pileup} systematic below), so the displacement-first choice, rather
than perfect pileup-track identification, is what stabilizes the reconstruction.

\textit{Timing observables and figure of merit.}---Each barrel-layer hit time is
first converted to an estimate of the track production time,
$t_i^{\rm SV}=t_i^{\rm hit}-L_i/(\beta_i c)$, where $L_i$ is the helix path length
from the reconstructed production point to the timing layer; in the baseline we
set $\beta_i=1$ for the charged hadrons, and the residual bias of this
approximation, which grows for the softest tracks, is varied over the
track-momentum-dependent hadron-mass envelope as a systematic on the timing
observables; its size as a function of track $p_T$ and its cancellation in
$\dTSV$ are quantified in the Supplemental Material (Fig.~\ref{fig:betabias}).
The start time $t_0$ is the mean production time of prompt
central reference tracks ($|d_0|<1\mm$, $|\eta|<1.48$, $p_T>0.7\GeV$); with
$\mathcal{O}(15)$ such tracks per event sharing one production time, the
statistical uncertainty is $\sigmat/\sqrt{N}\!\sim\!8\ps$. This is a statistical
floor only: correlated effects---beam-spot size, detector alignment, and
per-track resolution variations---do not scale as $1/\sqrt{N}$ and inflate the
realistic per-event start-time precision toward $\sigmat$ itself. We therefore
take $\simeq30\ps$ as the realistic start-time resolution throughout and quantify
its impact as the \emph{Start time} systematic below, where the quoted results
are shown to be insensitive to this choice. A usable reference is found in
$94\text{--}96\%$ of signal events (and in more of the high-multiplicity QCD
background). The few percent of events with no prompt reference cannot define
$t_0$ and therefore fail the 4D requirement, leaving the 3D baseline untouched.
The realistic hard-scatter resolution of the barrel layer is $\simeq30\ps$ at
start of life, rising to $\simeq60\ps$ at the end of operation~\cite{Addesa:2025kyl}; the long-$\ctau$ results, where the signal delay reaches hundreds
of ps, are insensitive to this value, whereas the short-$\ctau$ crossover region is
not.

\begin{figure}[t]
\centering
\includegraphics[width=\columnwidth]{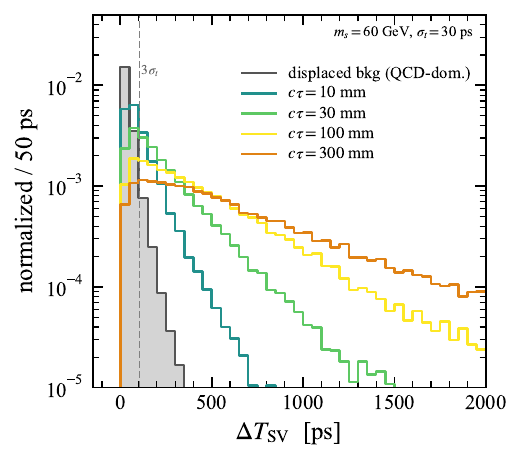}
\caption{Normalized vertex time delay $\dTSV$, constructed from path-length-corrected
SV-track production times after the 3D spatial selection, for the signal at
several decay lengths ($\ms=60\GeV$, $\sigmat=30\ps$) and for the displaced
background, which is dominated by QCD $b\bar bjj$. The background peaks at
$\dTSV\approx0$ (prompt heavy-flavour decays at $t_0$), whereas the signal shifts to
progressively larger delays as $\ctau$ increases. The dashed line indicates the nominal $3\sigmat$
threshold.}
\label{fig:dt}
\end{figure}

For the selected vertex we define the time delay
$\dTSV\equiv\langle t_i^{\rm SV}\rangle_{\rm SV}-t_0$, averaged over its timed
tracks, and the late-track fraction $\flate$, the fraction of timed vertex tracks
with $t_i^{\rm SV}$ later than $3\sigmat$ after $t_0$. Because the barrel layer
covers $|\eta|<1.48$, only the central tracks of a vertex carry a time measurement; across the
grid $80\text{--}87\%$ of spatially selected vertices retain at least one timed
track ($84.9\%$ at the benchmark $\ms=55\GeV$, $\ctau=100\mm$; see
Fig.~\ref{fig:timedsv} in the Supplemental Material), so $\dTSV$ is
defined for the large majority, and the minority with no timed track simply fail
the 4D
requirement. A genuine long-lived decay
produces a vertex whose tracks share a common positive delay set by the parent
flight time, whereas a prompt heavy-flavour vertex is centred at $t_0$; the two
observables capture, respectively, the magnitude and the coherence of the delay
(Fig.~\ref{fig:dt}). The nominal 4D selection requires $\dTSV>3\sigmat$ and
$\flate>0.3$ in addition to the spatial selection. The $\flate$ threshold is fixed
\emph{a priori} at $0.3$: it demands that at least roughly one third of the timed
tracks be late---enough to reject a single mistimed outlier while retaining the
majority of a coherent late vertex. The gain is insensitive to this choice,
changing by ${<}0.1\%$ over $\flate\in[0.2,0.6]$ (Table~\ref{tab:robust}). An
optimized working point instead applies $\dTSV>300\ps$, an
illustrative tighter requirement that trades signal efficiency for a much smaller
background tail.

Because absolute yields depend on detector and pileup modeling beyond the scope of a fast simulation, we quote a normalization-free figure of merit: the ratio of the timing-enhanced sensitivity to the spatial-only sensitivity,
\begin{equation}
  \Gain \;\equiv\;
  \frac{(S/\sqrt{B})_{\rm 4D}}{(S/\sqrt{B})_{\rm 3D}}
  \;=\; \frac{r_S}{\sqrt{r_B}}, \qquad
  r_{S,B} \equiv \frac{N^{\rm 4D}_{S,B}}{N^{\rm 3D}_{S,B}},
  \label{eq:gain}
\end{equation}
where $r_S$ and $r_B$ are the fractions of spatially selected signal and
background vertices that survive the timing requirement. The luminosity, the
absolute cross sections, and any common acceptance cancel in Eq.~(\ref{eq:gain}):
$\Gain>1$ indicates that the timing layer separates signal from background beyond
what spatial information alone can achieve, independent of normalization. It is
the robust deliverable of this work. Because $r_S$ and $r_B$ are constructed from
the same spatial vertices in the numerator and denominator, $\Gain$ is also
independent of the timing-layer acceptance entering the 3D baseline, which is
purely spatial. This 3D baseline applies the same displaced-vertex requirements
as the recent ATLAS search~\cite{ATLAS:2024grv}---a minimum track impact
parameter, a vertex mass-to-size ratio, and a track-$p_T$ sum---so that $\Gain$
isolates the effect of timing on a like-for-like spatial selection; it serves as
an internal reference for that purpose, not as a reproduction of the full
multivariate ATLAS result.

\textit{Backgrounds.}---The dominant irreducible background is the prompt
production of displaced-vertex candidates from heavy-flavour decays. We generate
three contributions, each normalized by its own generated cross section: QCD
$b\bar bjj$ ($540\pb$), electroweak $b\bar bjj$ ($0.21\pb$), and SM VBF
$h\to b\bar b$ ($0.17\pb$). After selection, the total is dominated by QCD
$b\bar bjj$, which supplies $99.9\%$ of the spatially selected background because
its cross section exceeds those of the other contributions by three orders of
magnitude. All three are time-prompt: their vertices arise from $b$- and
$c$-hadron decays at the hard-scatter time, populating $\dTSV\approx0$ (median
$25\ps$, median $\RSV=5\mm$). Because these vertices peak at $\dTSV\approx0$, the
late-time tail on which the timing selection acts is populated by the detector
timing response and by pileup, not by the tail of the QCD matrix element. The
leading-order simulation therefore determines this tail shape only
approximately; in a real analysis, it would be constrained by the $\dTSV<0$
sideband---which mirrors the prompt timing response---rather than by the
perturbative order of the calculation. This is the physical origin of the timing
discrimination: the background that survives the spatial selection is precisely
the population that the timing layer is designed to reject, and the QCD
continuum is even more prompt than SM $h\to b\bar b$ (timing survival $7\%$
versus $11\%$). Fully hadronic $t\bar t$ production yields a similar $b$-rich
multijet final state and can in principle populate the selection, but the
forward, high-mass VBF dijet requirement strongly suppresses its central
topology, leaving it subdominant to the QCD continuum and, being equally
time-prompt, subject to the same timing rejection. Other displaced or
instrumental backgrounds fall into two classes. Photon conversions, hadronic
material interactions, and fake or pileup-combinatorial vertices are prompt in
collision time and are suppressed, not enhanced, by the timing
requirement---with the random-late pileup component bracketed by the
\emph{Pileup} systematic below. $K^0_S$ and $\Lambda$ decays, by contrast, are
genuinely late: with proper decay lengths $c\tau=26.8$ and $79\mm$, they carry
true flight-time delays and would not be rejected by timing. They are instead
removed \emph{before} the timing step by the spatial selection, because a $V^0$
is a two-prong, low-mass object ($m_{K^0_S}=0.498$, $m_\Lambda=1.116\GeV$) that
fails both the $\geq3$-track requirement and the $m_{\rm SV}/\Delta R>4\GeV$
mass-to-size cut, whereas signal vertices carry masses of tens of GeV. For a
$V^0$ to enter the 3D selection, it would need to acquire additional tracks
accidentally and reach a vertex mass an order of magnitude above that of the
parent---a combinatorially negligible configuration---so we bound the $V^0$
contribution to the spatially selected background, and hence to $r_B$ and
$\Gain$, at the sub-percent level. A fast simulation cannot model these
instrumental components reliably; a real analysis controls them with material
vetoes, $V^0$ mass tags, and a data-driven background
estimate~\cite{ATLAS:2024grv,CMS:2021juv}. To first approximation, they enter as
a normalization or tail-shape uncertainty on the prompt background rather than
as a genuine late population---an assumption that only collision data can
validate through dedicated control regions: the $\dTSV<0$ sideband to calibrate
the prompt timing response and its tail, a low-$m_{\rm SV}/\Delta R$ or low-$\sum
p_T$ region to enrich prompt heavy-flavour vertices, a VBF sideband at lower
$m_{jj}$ or $|\Delta\eta_{jj}|$ to control the QCD continuum, and
random-crossing overlays to assess the pileup-combinatorial tail. Their
residual effect is confined to the absolute normalization, which we bracket
through $\kappa_b$, and does not alter the gain $\Gain$.

\begin{table}[t]
\centering
\caption{Robustness of the timing gain at the benchmark
($\ms=55\GeV$, $\ctau=100\mm$). Each block varies one nuisance parameter from the
baseline working point; $r_{S,B}$ are the timing-survival fractions of the
spatially selected signal and (cross-section-weighted) background, and
$\Gain=r_S/\sqrt{r_B}$. The nominal gain is stable or increases as the assumed
timing resolution entering the $3\sigmat$ significance window increases, with the
track-time smearing held at its start-of-life value, so the $\sigmat=30\ps$
baseline is the conservative choice. It varies by $\lesssim10\%$ across the
displacement cut. To disentangle the timing resolution from the
$\sigmat$-scaled threshold, we replace $\dTSV>3\sigmat$ with a fixed absolute
threshold $\dTSV>90\ps$ (keeping $\flate>0.3$): $\Gain$ is reproduced to within
$1\%$ at every $\sigmat$ ($2.8$, $3.2$, $4.6$, $5.7$ for $\sigmat=30$--$60\ps$).
With the absolute threshold fixed, the residual $\sigmat$ dependence enters
only through the coherence cut $\flate$, whose late-track boundary is itself
defined relative to $3\sigmat$; the near-invariance of $\Gain$ under this
replacement therefore shows that the background rejection at the nominal point is
driven by $\flate>0.3$---the requirement that the delay be coherent across the
vertex tracks---rather than by the absolute delay threshold. The
late-track-fraction requirement is itself flat: $\Gain$ changes by
${<}0.1\%$ for $\flate\in[0.2,0.6]$. The optimized
$\dTSV>300\ps$ point is based on $200$ raw QCD vertices ($\simeq7\%$ Monte Carlo
statistics); tighter thresholds exhaust the simulated tail and only bracket the
trend. Because $r_B$ is common to all masses, none of these variations shifts the
mass pivot of Fig.~\ref{fig:pivot}.}
\label{tab:robust}
\begin{ruledtabular}
\begin{tabular}{lccc}
Variation & $r_S$ & $r_B$ & $\Gain$ \\
\colrule
\multicolumn{4}{l}{\textit{Nominal point} ($\dTSV>3\sigmat$, $\flate>0.3$)}\\
Baseline ($\sigmat=30\ps$, $\RSV>3\mm$) & $0.755$ & $0.073$    & $2.8$ \\
$\sigmat=35\ps$                         & $0.731$ & $0.051$    & $3.2$ \\
$\sigmat=50\ps$                         & $0.664$ & $0.021$    & $4.6$ \\
$\sigmat=60\ps$                         & $0.620$ & $0.012$    & $5.7$ \\
$\RSV>2\mm$                             & $0.750$ & $0.063$    & $3.0$ \\
$\RSV>5\mm$                             & $0.769$ & $0.101$    & $2.4$ \\
\colrule
\multicolumn{4}{l}{\textit{Optimized point} ($\dTSV>300\ps$)}\\
Baseline                                & $0.464$ & $2.1\times10^{-3}$ & $10.2$ \\
$\dTSV>150\ps$                          & $0.664$ & $2.1\times10^{-2}$ & $4.6$  \\
$\dTSV>500\ps$ (MC-stat $\sim20\%$)     & $0.276$ & $2.7\times10^{-4}$ & $16.8$ \\
\end{tabular}
\end{ruledtabular}
\end{table}

\textit{Results.}---The timing layer acts almost exclusively on the background
(Table~\ref{tab:cutflow} in the Supplemental Material): across the grid, it
retains $r_B=0.073$ of the spatially selected background at the nominal point
and $r_B=2.1\times10^{-3}$ at the optimized point, while retaining
$60\text{--}80\%$ of the signal for $\ctau\gtrsim30\mm$. Through
Eq.~(\ref{eq:gain}), this yields a gain of $\Gain=2.8$ at the nominal working
point and $\Gain\simeq10$ at the optimized point for the benchmark
$\ms=60\GeV$, $\ctau=100\mm$, rising to $\Gain\simeq15$ at $\ctau=1000\mm$,
where essentially all of the signal is delayed but the prompt background is
fully rejected. For $\ctau\lesssim10\mm$, the gain falls below unity: the
signal delay becomes comparable to $\sigmat$, so the timing cut discards signal
without a compensating reduction in background, and the spatial selection alone
is preferable.

\begin{figure}[t]
\centering
\includegraphics[width=\columnwidth]{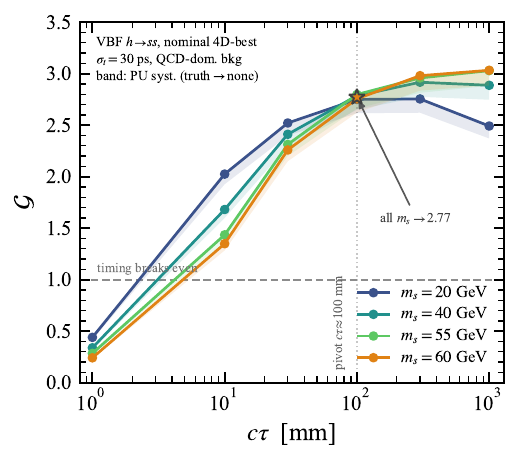}
\caption{Normalization-free gain $\Gain$ [Eq.~(\ref{eq:gain})] as a function of decay
length for the four scalar masses at the nominal 4D working point
($\sigmat=30\ps$, QCD-dominated background). The dashed line indicates the
break-even point, $\Gain=1$. At $\ctau\simeq100\mm$, all masses converge to
$\Gain\simeq2.8$ (star), and the mass ordering reverses: lighter scalars yield a
larger gain below the pivot, whereas heavier scalars do so above it. Because $r_B$ is common to
all masses, the background normalization cancels in $\Gain$, and the location of
the pivot is determined by the signal kinematics alone; threshold and tail-shape
systematics are examined separately in the text.}
\label{fig:pivot}
\end{figure}

The dependence on scalar mass, shown in Fig.~\ref{fig:pivot}, is the central
structural result. As anticipated from the kinematics---the lab-frame decay
length is $L_{\rm lab}\simeq\beta\gamma\,\ctau$ with $\beta\gamma\simeq p_s/\ms$,
so the mean boost scales as $\langle\gamma_s\rangle\propto1/\ms$ and lighter
scalars travel farther for a given proper lifetime---the two mass regimes trade
places at a well-defined decay length. At short $\ctau$, the larger boost of the
light scalar pushes more decays beyond the $3\mm$ displacement cut and increases
their time delay, so the lighter mass yields the larger gain. At long $\ctau$,
the same boost carries the light scalar out of the tracker, reducing the signal,
and the heavier, slower scalar prevails. The two effects balance at
$\ctau\simeq100\mm$, where all four masses collapse to $\Gain\simeq2.8$ and the
ordering inverts---a crossover that is invisible to a purely spatial analysis, in
which the heavier mass is favored at every lifetime, and that emerges only once
the temporal dimension is included. This pivot at $\ctau\simeq100\mm$ is specific
to the assumed tracker fiducial radius and barrel acceptance and would shift for
a different detector geometry.

\textit{Projected reach.}---To place the gain in physical context, we translate
the selection into an illustrative, statistics-limited 95\% CL upper limit on
$\mathrm{BR}(h\to ss)$ at $\mathcal{L}=3000\fbinv$ (again assuming
$\mathrm{BR}(s\to b\bar b)=1$), using
the asymptotic $\mathrm{CL}_s$
procedure~\cite{Cowan:2010js,Read:2002hq} for a single-bin counting experiment;
the signal strength scales linearly with the branching ratio. These limits are an
idealized floor: the absolute background normalization is not predicted by the
fast simulation and must be supplied by a data-driven estimate in a real
experiment, an effect that we bracket below through $\kappa_b$.
Figure~\ref{fig:limit}
in the Supplemental Material shows the expected limit for $\ms=60\GeV$. Adding
timing tightens the floor from $\mathrm{BR}\simeq0.19$ (3D) to $\simeq0.019$
(optimized 4D) at $\ctau=100\mm$, with the improvement tracking the gain of
Fig.~\ref{fig:pivot}; the best reach across the plane,
$\mathrm{BR}\simeq0.019$, is obtained for $\ms=55\text{--}60\GeV$ at
$\ctau=100\mm$. The upper point, $\ms=60\GeV$, lies close to the
$\ms\to m_h/2$ kinematic threshold, where the narrow phase space makes the decay
kinematics more sensitive to off-shell and generator effects, so we treat it as a
near-threshold benchmark rather than a sharp optimum.

The complementarity to existing searches is sharpest at long lifetimes. The
spatial search of Ref.~\cite{ATLAS:2024grv} is most sensitive around
$\ctau\simeq5\text{--}70\mm$ and loses reach beyond $\sim100\mm$, as decays
increasingly fall outside the tracker fiducial volume and prompt-background
rejection no longer improves. By contrast, the timing gain grows with $\ctau$,
since the exploited delay scales with the flight distance. Although the in-tracker
signal acceptance falls for both approaches, the 4D limit degrades far more
gently and remains at $\mathrm{BR}\simeq0.02\text{--}0.04$ out to
$\ctau=1000\mm$, the regime in which a purely spatial limit has risen above the
physical $\mathrm{BR}=1$ boundary. Timing therefore extends the reach into the
long-lifetime region that a spatial-only selection cannot cover.

\textit{Systematic uncertainties.}---Four effects bound the projection, and
Table~\ref{tab:robust} summarizes the stability of the gain $\Gain$ at the
benchmark against the principal detector and selection nuisances.
\emph{Pileup.} The signal is robust against how pileup (pu) tracks are removed:
switching from the truth-association proxy (pu${=}$truth) to no removal at all
(pu${=}$none) shifts the signal yield and vertex-matching purity by $\lesssim3\%$,
because a displaced vertex is a coherent object---several tracks sharing a common
late time and a common displaced position---that random pileup cannot mimic. The
background is not. With pileup tracks left in, the $\pm180\ps$ pileup time spread
is attached to the prompt QCD vertices and smears them into the late tail,
raising the background timing survival by $\simeq10\%$ at the nominal point and
more than doubling it (a factor of $\simeq2.4$) at the optimized one in this
no-association extreme. This brackets the gain between its ideal-association
value and a conservative no-association value, degrading $\Gain$ by
$\lesssim5\%$ (nominal) and $\lesssim35\%$ (optimized). These figures are
evaluated directly for the barrel timing layer (QCD re-simulated with no
pileup-track association, $2.2\times10^{6}$ events), rather than transferred from
a different timing geometry. The realistic case lies between the two and close to
the ideal, because the experiment associates tracks to vertices by time
\emph{and} position: a per-track time veto cannot be used to remove pileup here,
since it would also discard the genuinely late signal. Separating coherent late
signal from random late pileup is therefore intrinsically a vertex-level rather
than a track-level operation. Crucially, $r_B$ is common to every signal point,
so the mass-trend reversal of Fig.~\ref{fig:pivot} remains intact under this
uncertainty.
\emph{Start time.} Degrading $t_0$ from its idealized $\simeq8\ps$ to a realistic
$30\ps$ leaves the long-lifetime gain unchanged, because the signal delay there
($\dTSV\gtrsim400\ps$) far exceeds $\sigma_{t_0}$. Even in the short-lifetime
region, the effect is modest: the broader start time smears the prompt background
by the same amount, raising the nominal background survival by $\simeq16\%$ and
lowering the nominal-point gain by $\lesssim7\%$ for $\ctau=10$--$30\mm$, while
shifting the break-even point by $\lesssim0.3\mm$. The optimized
$\dTSV>300\ps$ working point is essentially immune ($\lesssim2\%$ change in
background survival), since a $30\ps$ resolution cannot promote a prompt vertex
beyond a $300\ps$ threshold.
\emph{Tail statistics.} The optimized $\dTSV>300\ps$ working point retains
$\simeq200$ QCD vertices out of the $2.2\times10^{6}$-event sample, corresponding
to a $\simeq7\%$ Monte Carlo statistical uncertainty on $r_B$ that propagates
directly to the optimized gain and to the corresponding limit. The shape of the
prompt late-time tail beyond the simulated statistics is not predicted by a fast
simulation and would, in a real analysis, be constrained by a sideband fit to the
$\dTSV$ distribution. We therefore quote this working point as an illustrative
tighter selection rather than a fully exploited optimum.
\emph{Background normalization.} As a phenomenological projection, this study has
no data with which to normalize the QCD continuum, which yields
$\mathcal{O}(10^{6}\text{--}10^{8})$ selected events at $3000\fbinv$; a
fractional background uncertainty of even $\kappa_b\sim1\%$ would then exceed the
statistical error by one to two orders of magnitude. The absolute limits are
therefore a statistics-limited floor, and we present them as a function of an
assumed $\kappa_b$ that brackets what a data-driven control-region estimate would
achieve in a real experiment. This systematics-dominated regime in fact
strengthens the case for timing: because the optimized selection reduces the
background by a further factor of $\simeq500$, the 4D limit is far less
sensitive to $\kappa_b$ than the 3D one. At $\kappa_b=1\%$, the spatial search no
longer excludes any physically allowed branching ratio, with its limit lying
above unity over the entire lifetime range, whereas the optimized 4D search still
excludes $\mathrm{BR}\lesssim0.1$ for $\ctau\gtrsim10\mm$---a median limit lower
by a factor of $\simeq216$ at the benchmark $\ctau=100\mm$
(Fig.~\ref{fig:limit_kappa} in the Supplemental Material). The
advantage of timing over a purely spatial search therefore grows as the
background uncertainty increases. Moreover, a fast simulation does not capture the
degradation of displaced-track efficiency and impact-parameter resolution with
displacement, nor displaced-track fakes. These act on the spatially selected
vertices that enter the numerator and denominator of both $r_S$ and $r_B$ in a
correlated way and therefore largely cancel in the ratio $\Gain$, but they would
have to be calibrated from data or with a full simulation before the absolute
limits could be regarded as experiment-grade. None of these effects affects the
normalization-free gain $\Gain$.

\textit{Conclusions.}---We have shown that the forward tag jets of vector-boson
fusion identify a hard-scatter vertex whose prompt central tracks, timed by a
barrel timing layer, supply a clean event-by-event start time that converts a
displaced vertex into a four-dimensional object, and that the resulting timing
layer acts almost exclusively on the prompt heavy-flavor background that no
spatial cut can remove. The mechanism is internally consistent: the same forward,
high-mass dijet that defines the VBF signal region also tags the vertex
furnishing the timing reference, so the temporal handle comes for free with the
spatial one.

Quantitatively, timing improves the sensitivity by a normalization-free factor
$\Gain=2.8$ at the nominal working point and by up to an order of magnitude for a
tighter delayed-vertex requirement. Under the explicit assumption
$\mathrm{BR}(s\to b\bar b)=1$, this corresponds to an illustrative,
statistics-limited $\mathrm{BR}(h\to ss)$ floor that can approach $\simeq2\%$ for
$\ms=55\text{--}60\GeV$ at intermediate lifetimes. Without the timing
dimension, separating a genuine long-lived decay from the prompt heavy-flavor
continuum requires a spatial displacement large enough to overcome that
continuum's sub-millimeter tail; with timing included, the same separation is
achieved through a time delay that the background cannot produce, and the two
requirements are independent. The structural signature of this independence is
the mass-trend reversal at $\ctau\simeq100\mm$, where the boost-driven and
containment-driven mass dependences cancel.

These projections are complementary to recent spatial
searches~\cite{ATLAS:2024grv,CMS:2021juv,CMS:2024lowmass}, one of which already
constrains this VBF topology with Run-2 data but is purely spatial and, in its
VBF region, systematics-limited. At the HL-LHC, additional luminosity alone
cannot improve a background-systematics-limited spatial search; the timing layer
reduces the surviving background by a further factor of $\simeq500$,
substantially reducing the dependence on the background normalization, and its
advantage is largest precisely at long lifetimes, $\ctau\gtrsim100\mm$, where the
spatial handle weakens. Timing is thus the lever that keeps this search class
improving where luminosity and spatial information alone have saturated.

What remains is specific to an experimental realization rather than to the
method: a data-driven control-region estimate of the QCD continuum, which only
collision data can provide and whose impact we have bracketed through $\kappa_b$,
and a dedicated mapping of the start-time and pileup systematics across the
short-lifetime region for the barrel-only timing configuration. Neither affects
the gain $\Gain$, which we therefore advance as the primary, detector-robust
result.

\begin{acknowledgments}
The author thanks colleagues at the Institute of High Energy
Physics for stimulating discussions.
This work is supported by the Internal Research Fund of the
Institute of High Energy Physics,
Chinese Academy of Sciences.
\end{acknowledgments}

\bibliographystyle{apsrev4-2}
\bibliography{reference}

@article{Alwall:2014hca,
    author = "Alwall, J. and Frederix, R. and Frixione, S. and Hirschi, V. and Maltoni, F. and Mattelaer, O. and Shao, H. -S. and Stelzer, T. and Torrielli, P. and Zaro, M.",
    title = "{The automated computation of tree-level and next-to-leading order differential cross sections, and their matching to parton shower simulations}",
    eprint = "1405.0301",
    archivePrefix = "arXiv",
    primaryClass = "hep-ph",
    reportNumber = "CERN-PH-TH-2014-064, CP3-14-18, LPN14-066, MCNET-14-09, ZU-TH-14-14",
    doi = "10.1007/JHEP07(2014)079",
    journal = "JHEP",
    volume = "07",
    pages = "079",
    year = "2014"
}

@article{Bierlich:2022pfr,
    author = "Bierlich, Christian and others",
    title = "{A comprehensive guide to the physics and usage of PYTHIA 8.3}",
    eprint = "2203.11601",
    archivePrefix = "arXiv",
    primaryClass = "hep-ph",
    reportNumber = "LU-TP 22-16, MCNET-22-04, FERMILAB-PUB-22-227-SCD",
    doi = "10.21468/SciPostPhysCodeb.8",
    journal = "SciPost Phys. Codeb.",
    volume = "2022",
    pages = "8",
    year = "2022"
}

@article{deFavereau:2013fsa,
    author = "de Favereau, J. and Delaere, C. and Demin, P. and Giammanco, A. and Lema{\^\i}tre, V. and Mertens, A. and Selvaggi, M.",
    collaboration = "DELPHES 3",
    title = "{DELPHES 3, A modular framework for fast simulation of a generic collider experiment}",
    eprint = "1307.6346",
    archivePrefix = "arXiv",
    primaryClass = "hep-ex",
    doi = "10.1007/JHEP02(2014)057",
    journal = "JHEP",
    volume = "02",
    pages = "057",
    year = "2014"
}

@article{Cacciari:2008gp,
    author = "Cacciari, Matteo and Salam, Gavin P. and Soyez, Gregory",
    title = "{The anti-$k_t$ jet clustering algorithm}",
    eprint = "0802.1189",
    archivePrefix = "arXiv",
    primaryClass = "hep-ph",
    reportNumber = "LPTHE-07-03",
    doi = "10.1088/1126-6708/2008/04/063",
    journal = "JHEP",
    volume = "04",
    pages = "063",
    year = "2008"
}

@article{Cacciari:2011ma,
    author = "Cacciari, Matteo and Salam, Gavin P. and Soyez, Gregory",
    title = "{FastJet User Manual}",
    eprint = "1111.6097",
    archivePrefix = "arXiv",
    primaryClass = "hep-ph",
    reportNumber = "CERN-PH-TH-2011-297",
    doi = "10.1140/epjc/s10052-012-1896-2",
    journal = "Eur. Phys. J. C",
    volume = "72",
    pages = "1896",
    year = "2012"
}

@article{PDF4LHCWorkingGroup:2022cjn,
    author = "Ball, Richard D. and others",
    collaboration = "PDF4LHC Working Group",
    title = "{The PDF4LHC21 combination of global PDF fits for the LHC Run III}",
    eprint = "2203.05506",
    archivePrefix = "arXiv",
    primaryClass = "hep-ph",
    reportNumber = "Edinburgh 2021/31, FERMILAB-PUB-22-121-QIS-SCD-T, MSUHEP-22-010, SMU-HEP-22-01, Nikhef 2021-033",
    doi = "10.1088/1361-6471/ac7216",
    journal = "J. Phys. G",
    volume = "49",
    number = "8",
    pages = "080501",
    year = "2022"
}

@article{Liu:2018wte,
    author = "Liu, Jia and Liu, Zhen and Wang, Lian-Tao",
    title = "{Enhancing Long-Lived Particles Searches at the LHC with Precision Timing Information}",
    eprint = "1805.05957",
    archivePrefix = "arXiv",
    primaryClass = "hep-ph",
    reportNumber = "FERMILAB-PUB-18-173-T, EFI-18-7",
    doi = "10.1103/PhysRevLett.122.131801",
    journal = "Phys. Rev. Lett.",
    volume = "122",
    number = "13",
    pages = "131801",
    year = "2019"
}

@article{ATLAS:2024grv,
    author = "{ATLAS Collaboration}",
    title = "{Search for Light Long-Lived Particles in $pp$ Collisions at $\sqrt{s}=13$ TeV Using Displaced Vertices in the ATLAS Inner Detector}",
    eprint = "2403.15332",
    archivePrefix = "arXiv",
    primaryClass = "hep-ex",
    reportNumber = "CERN-EP-2024-086",
    doi = "10.1103/PhysRevLett.133.161803",
    journal = "Phys. Rev. Lett.",
    volume = "133",
    number = "16",
    pages = "161803",
    year = "2024",
    note = "[Erratum: Phys.Rev.Lett. 135, 259901 (2025)]"
}

@article{Cowan:2010js,
    author = "Cowan, Glen and Cranmer, Kyle and Gross, Eilam and Vitells, Ofer",
    title = "{Asymptotic formulae for likelihood-based tests of new physics}",
    eprint = "1007.1727",
    archivePrefix = "arXiv",
    primaryClass = "physics.data-an",
    doi = "10.1140/epjc/s10052-011-1554-0",
    journal = "Eur. Phys. J. C",
    volume = "71",
    pages = "1554",
    year = "2011",
    note = "[Erratum: Eur.Phys.J.C 73, 2501 (2013)]"
}

@article{Read:2002hq,
    author = "Read, Alexander L.",
    editor = "Whalley, M. R. and Lyons, L.",
    title = "{Presentation of search results: The $CL_s$ technique}",
    doi = "10.1088/0954-3899/28/10/313",
    journal = "J. Phys. G",
    volume = "28",
    pages = "2693--2704",
    year = "2002"
}

@techreport{CERN-LHCC-2020-007,
      author        = "{ATLAS Collaboration}",
      title         = "{Technical Design Report:  A High-Granularity Timing
                       Detector for the ATLAS Phase-II Upgrade}",
      institution   = "CERN",
      reportNumber  = "CERN-LHCC-2020-007, ATLAS-TDR-031",
      address       = "Geneva",
      year          = "2020",
      url           = "https://cds.cern.ch/record/2719855",
}

@techreport{CMS-TDR-020,
      author        = "{CMS Collaboration}",
      title         = "{A MIP Timing Detector for the CMS Phase-2 Upgrade}",
      institution   = "CERN",
      reportNumber  = "CERN-LHCC-2019-003, CMS-TDR-020",
      address       = "Geneva",
      year          = "2019",
      url           = "https://cds.cern.ch/record/2667167",
}

@article{Strassler:2006im,
    author = "Strassler, Matthew J. and Zurek, Kathryn M.",
    title = "{Echoes of a hidden valley at hadron colliders}",
    eprint = "hep-ph/0604261",
    archivePrefix = "arXiv",
    primaryClass = "hep-ph",
    doi = "10.1016/j.physletb.2007.06.055",
    journal = "Phys. Lett. B",
    volume = "651",
    pages = "374--379",
    year = "2007"
}

@article{Strassler:2006ri,
    author = "Strassler, Matthew J. and Zurek, Kathryn M.",
    title = "{Discovering the Higgs through highly-displaced vertices}",
    eprint = "hep-ph/0605193",
    archivePrefix = "arXiv",
    primaryClass = "hep-ph",
    doi = "10.1016/j.physletb.2008.02.008",
    journal = "Phys. Lett. B",
    volume = "661",
    pages = "263--267",
    year = "2008"
}

@article{Curtin:2013fra,
    author = "Curtin, David and others",
    title = "{Exotic decays of the 125 GeV Higgs boson}",
    eprint = "1312.4992",
    archivePrefix = "arXiv",
    primaryClass = "hep-ph",
    doi = "10.1103/PhysRevD.90.075004",
    journal = "Phys. Rev. D",
    volume = "90",
    number = "7",
    pages = "075004",
    year = "2014"
}

@article{Alimena:2019zor,
    author = "Alimena, Juliette and others",
    title = "{Searching for long-lived particles beyond the Standard Model at the Large Hadron Collider}",
    eprint = "1903.04497",
    archivePrefix = "arXiv",
    primaryClass = "hep-ph",
    doi = "10.1088/1361-6471/ab4574",
    journal = "J. Phys. G",
    volume = "47",
    number = "9",
    pages = "090501",
    year = "2020"
}

@article{ATLAS:2021jig,
    author = "{ATLAS Collaboration}",
    title = "{Search for exotic decays of the Higgs boson into long-lived particles in $pp$ collisions at $\sqrt{s}=13$ TeV using displaced vertices in the ATLAS inner detector}",
    eprint = "2107.06092",
    archivePrefix = "arXiv",
    primaryClass = "hep-ex",
    reportNumber = "CERN-EP-2021-098",
    doi = "10.1007/JHEP11(2021)229",
    journal = "JHEP",
    volume = "11",
    pages = "229",
    year = "2021"
}

@article{CMS:2024lowmass,
  author        = "{CMS Collaboration}",
  title         = "{Search for light long-lived particles decaying to displaced jets in proton-proton collisions at $\sqrt{s}=13.6$ TeV}",
  eprint        = "2409.10806",
  archivePrefix = "arXiv",
  primaryClass  = "hep-ex",
  journal       = "Rept. Prog. Phys.",
  volume        = "88",
  number        = "3",
  pages         = "037801",
  year          = "2025",
  doi           = "10.1088/1361-6633/adaa13"
}

@article{CMS:2021juv,
  author        = "{CMS Collaboration}",
  title         = "{Search for long-lived particles using displaced jets in proton-proton collisions at $\sqrt{s}=13$ TeV}",
  eprint        = "2012.01581",
  archivePrefix = "arXiv",
  primaryClass  = "hep-ex",
  journal       = "Phys. Rev. D",
  volume        = "104",
  number        = "1",
  pages         = "012015",
  year          = "2021",
  doi           = "10.1103/PhysRevD.104.012015"
}

@inproceedings{Cerri:2018skj,
    author = "Cerri, Olmo",
    collaboration = "CMS",
    title = "{CMS precision timing physics impact for the HL-LHC upgrade}",
    booktitle = "{13th Conference on the Intersections of Particle and Nuclear Physics}",
    eprint = "1810.00860",
    archivePrefix = "arXiv",
    primaryClass = "physics.ins-det",
    reportNumber = "CIPANP2018-Cerri",
    month = "10",
    year = "2018"
}

@article{Flowers:2019eoc,
  author        = "Flowers, Zachary and Kang, Dong Woo and Meier, Quinn and Park, Seong Chan and Rogan, Christopher",
  title         = "{Timing information at HL-LHC: complete determination of masses of dark matter and long lived particle}",
  eprint        = "1903.05825",
  archivePrefix = "arXiv",
  primaryClass  = "hep-ph",
  journal       = "JHEP",
  volume        = "03",
  pages         = "132",
  year          = "2020",
  doi           = "10.1007/JHEP03(2020)132"
}

@article{Bhattacherjee:2021rml,
  author        = "Bhattacherjee, Biplob and Ghosh, Tapasi and Sengupta, Rhitaja and Solanki, Prabhat",
  title         = "{Dedicated triggers for displaced jets using timing information from electromagnetic calorimeter at HL-LHC}",
  eprint        = "2112.04518",
  archivePrefix = "arXiv",
  primaryClass  = "hep-ph",
  journal       = "JHEP",
  volume        = "08",
  pages         = "254",
  year          = "2022",
  doi           = "10.1007/JHEP08(2022)254"
}

@article{Banerjee:2019ktv,
  author        = "Banerjee, Shankha and Bhattacherjee, Biplob and Goudelis, Andreas and Herrmann, Bjorn and Sengupta, Dipan and Sengupta, Rhitaja",
  title         = "{Determining the lifetime of long-lived particles at the HL-LHC}",
  eprint        = "1912.06669",
  archivePrefix = "arXiv",
  primaryClass  = "hep-ph",
  journal       = "Eur. Phys. J. C",
  volume        = "81",
  pages         = "172",
  year          = "2021"
}

@article{Jones-Perez:2019plk,
  author        = "Jones-Perez, J. and Masias, J. and Ruiz-Alvarez, J. D.",
  title         = "{Search for Long-Lived Heavy Neutrinos at the LHC with a VBF Trigger}",
  eprint        = "1912.08206",
  archivePrefix = "arXiv",
  primaryClass  = "hep-ph",
  journal       = "Eur. Phys. J. C",
  volume        = "80",
  pages         = "642",
  year          = "2020"
}

@article{Bhattacherjee:2021zvv,
  author        = "Bhattacherjee, Biplob and Matsumoto, Shigeki and Sengupta, Rhitaja",
  title         = "{Long-Lived Light Mediators from Higgs boson Decay at HL-LHC, FCC-hh and a Proposal of Dedicated LLP Detectors for FCC-hh}",
  eprint        = "2111.02437",
  archivePrefix = "arXiv",
  primaryClass  = "hep-ph",
  journal       = "Phys. Rev. D",
  volume        = "106",
  number        = "9",
  pages         = "095018",
  year          = "2022",
  doi           = "10.1103/PhysRevD.106.095018"
}

@article{Acosta:2021qpx,
    author = "Acosta, Darin and others",
    editor = "Alimena, Juliette and others",
    title = "{Review of opportunities for new long-lived particle triggers in Run 3 of the Large Hadron Collider}",
    eprint = "2110.14675",
    archivePrefix = "arXiv",
    primaryClass = "hep-ex",
    reportNumber = "CERN-LPCC-2021-01",
    journal       = "",
    month = "10",
    year = "2021"
}

@article{deFlorian:2016spz,
  author        = "de Florian, D. and others",
  title         = "{Handbook of LHC Higgs Cross Sections: 4. Deciphering the Nature of the Higgs Sector}",
  eprint        = "1610.07922",
  archivePrefix = "arXiv",
  primaryClass  = "hep-ph",
  reportNumber  = "CERN-2017-002-M",
  doi           = "10.23731/CYRM-2017-002",
  journal       = "",
  year          = "2016"
}

@article{Cerri:2018rkm,
    author = "Cerri, O. and Xie, S. and Pena, C. and Spiropulu, M.",
    title = "{Identification of Long-lived Charged Particles using Time-Of-Flight Systems at the Upgraded LHC detectors}",
    eprint = "1807.05453",
    archivePrefix = "arXiv",
    primaryClass = "hep-ex",
    reportNumber = "FERMILAB-PUB-18-376-V",
    doi = "10.1007/JHEP04(2019)037",
    journal = "JHEP",
    volume = "04",
    pages = "037",
    year = "2019"
}

@article{Addesa:2025kyl,
    author = "Addesa, F. and others",
    title = "{The CMS barrel timing layer: test beam confirmation of module timing performance}",
    eprint = "2504.11209",
    archivePrefix = "arXiv",
    primaryClass = "physics.ins-det",
    reportNumber = "issn: 0168-9002",
    doi = "10.1016/j.nima.2025.170823",
    journal = "Nucl. Instrum. Meth. A",
    volume = "1081",
    pages = "170823",
    year = "2026"
}

@article{Cepeda:2019klc,
    author = "Cepeda, M. and others",
    editor = "Dainese, Andrea and Mangano, Michelangelo and Meyer, Andreas B. and Nisati, Aleandro and Salam, Gavin and Vesterinen, Mika Anton",
    title = "{Report from Working Group 2}: {Higgs Physics at the HL-LHC and HE-LHC}",
    eprint = "1902.00134",
    archivePrefix = "arXiv",
    primaryClass = "hep-ph",
    reportNumber = "CERN-LPCC-2018-04",
    doi = "10.23731/CYRM-2019-007.221",
    journal = "CERN Yellow Rep. Monogr.",
    volume = "7",
    pages = "221--584",
    year = "2019"
}

@article{Lee:2018pag,
  author        = {Lee, Lawrence and Ohm, Christian and Soffer, Abner and Yu, Tien-Tien},
  title         = {{Collider Searches for Long-Lived Particles Beyond the Standard Model}},
  eprint        = {1810.12602},
  archivePrefix = {arXiv},
  primaryClass  = {hep-ph},
  doi           = {10.1016/j.ppnp.2019.02.006},
  journal       = {Prog. Part. Nucl. Phys.},
  volume        = {106},
  pages         = {210--255},
  year          = {2019}
}

@techreport{CMS:Phase2TrackerTDR,
  collaboration = {CMS},
  title         = {{The Phase-2 Upgrade of the CMS Tracker}},
  institution   = {CERN},
  reportNumber  = {CERN-LHCC-2017-009, CMS-TDR-014},
  address       = {Geneva},
  year          = {2017},
  doi           = {10.17181/CERN.QZ28.FLHW},
  url           = {https://cds.cern.ch/record/2272264}
}

@techreport{CMS:Phase2L1TriggerTDR,
  collaboration = {CMS},
  title         = {{The Phase-2 Upgrade of the CMS Level-1 Trigger}},
  institution   = {CERN},
  reportNumber  = {CERN-LHCC-2020-004, CMS-TDR-021},
  address       = {Geneva},
  year          = {2020},
  note          = {Final version},
  url           = {https://cds.cern.ch/record/2714892}
}

@inproceedings{CMS:DisplacedVertexTrackTriggerPhase2L1,
  author        = {McCarthy, Ryan Edward},
  collaboration = {CMS},
  title         = {{Displaced Vertex Track Trigger for the CMS Phase-2 Level-1 Trigger Upgrade}},
  booktitle     = {{12th Large Hadron Collider Physics Conference}},
  journal       = {PoS},
  volume        = {LHCP2024},
  pages         = {274},
  year          = {2025},
  doi           = {10.22323/1.478.0274},
  reportNumber  = {CMS-CR-2024-313}
}

@article{Craig:2015pha,
  author        = {Craig, Nathaniel and Katz, Andrey and Strassler, Matt and Sundrum, Raman},
  title         = {{Naturalness in the Dark at the LHC}},
  eprint        = {1501.05310},
  archivePrefix = {arXiv},
  primaryClass  = {hep-ph},
  doi           = {10.1007/JHEP07(2015)105},
  journal       = {JHEP},
  volume        = {07},
  pages         = {105},
  year          = {2015}
}

@article{Curtin:2015fna,
  author        = {Curtin, David and Verhaaren, Christopher B.},
  title         = {{Discovering Uncolored Naturalness in Exotic Higgs Decays}},
  eprint        = {1506.06141},
  archivePrefix = {arXiv},
  primaryClass  = {hep-ph},
  doi           = {10.1007/JHEP12(2015)072},
  journal       = {JHEP},
  volume        = {12},
  pages         = {072},
  year          = {2015}
}

@article{Bertolini:2014bba,
  author        = {Bertolini, Daniele and Harris, Philip and Low, Matthew and Tran, Nhan},
  title         = {{Pileup Per Particle Identification}},
  eprint        = {1407.6013},
  archivePrefix = {arXiv},
  primaryClass  = {hep-ph},
  doi           = {10.1007/JHEP10(2014)059},
  journal       = {JHEP},
  volume        = {10},
  pages         = {059},
  year          = {2014}
}

@article{CMS:2014pgm,
  collaboration = {CMS},
  title         = {{Description and performance of track and primary-vertex reconstruction with the CMS tracker}},
  eprint        = {1405.6569},
  archivePrefix = {arXiv},
  primaryClass  = {physics.ins-det},
  doi           = {10.1088/1748-0221/9/10/P10009},
  journal       = {JINST},
  volume        = {9},
  pages         = {P10009},
  year          = {2014}
}

@article{Gligorov:2026review,
  author        = {Gligorov, Vladimir V.},
  title         = {{Towards triggerless four-dimensional detectors for High Energy Physics collider experiments}},
  eprint        = {2607.01496},
  archivePrefix = {arXiv},
  journal       = {},
  primaryClass  = {physics.ins-det},
  year          = {2026}
}

\clearpage
\twocolumngrid
\begin{center}
\textbf{\large Supplemental Material}\\[3pt]
\end{center}
\vspace{4pt}
\setcounter{figure}{0}
\renewcommand{\thefigure}{S\arabic{figure}}
\setcounter{table}{0}
\renewcommand{\thetable}{S\arabic{table}}

\begin{figure}[h]
\centering
\includegraphics[width=0.9\linewidth]{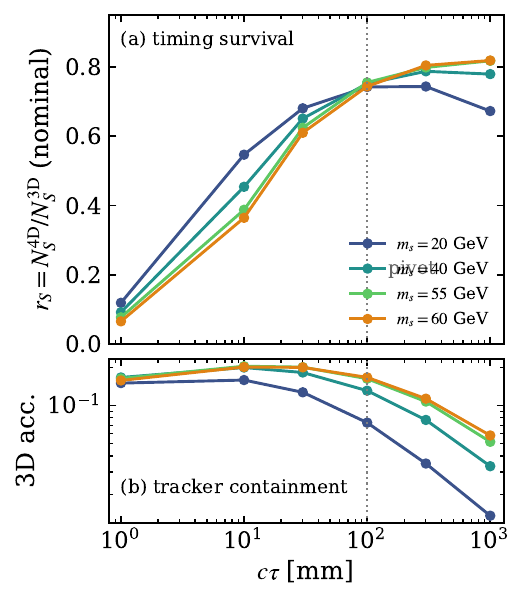}
\caption{(a) Nominal 4D signal timing survival, $r_S=N^{\rm 4D}_S/N^{\rm 3D}_S$, as a function of decay length for the four scalar masses; the curves converge at $\ctau\simeq100\mm$ (dotted line), where the mass ordering of the gain reverses. (b) Spatial (3D) acceptance, $N^{\rm 3D}_S/N_{\rm gen}$, which decreases most rapidly for the lightest, most boosted scalar at long $\ctau$. Since $r_B$ is common to all masses, panel (a) shows the full mass dependence of $\Gain$.}
\label{fig:rsdecomp}
\end{figure}

\begin{figure}[h]
\centering
\includegraphics[width=0.9\linewidth]{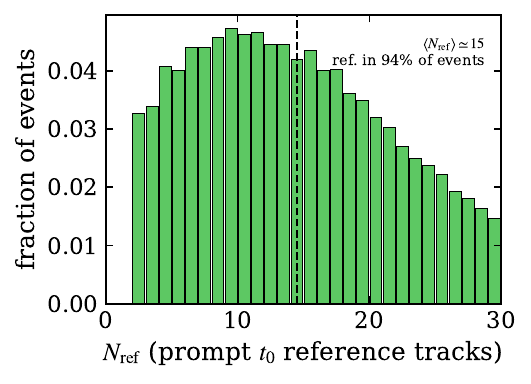}
\caption{Multiplicity of prompt reference tracks, $N_{\rm ref}$, used to determine $t_0$
at the benchmark point ($\ms=55\GeV$, $\ctau=100\mm$); the dashed line indicates the mean.
A usable reference is identified in $94\text{--}96\%$ of events across the grid.}
\label{fig:nref}
\end{figure}

\noindent
This supplement presents the selection cutflow (Table~\ref{tab:cutflow}), the
statistical procedure used to derive the projected limits, and the robustness studies
referenced in the main text: the timed-SV efficiency underlying the
barrel-timing configuration (Fig.~\ref{fig:timedsv}), the mass decomposition of
the timing survival that yields the gain pivot (Fig.~\ref{fig:rsdecomp}), the
start-time reference multiplicity (Fig.~\ref{fig:nref}), and the
illustrative branching-ratio limit together with its dependence on the assumed
background-normalization uncertainty $\kappa_b$
(Figs.~\ref{fig:limit} and \ref{fig:limit_kappa}). All results use the same
$2.2\times10^{6}$-event QCD sample and the full signal grid
($\ms\in\{20,40,55,60\}\GeV$, $\ctau\in\{1,10,30,100,300,1000\}\mm$) as in the main
analysis, with the barrel timing layer ($|\eta|<1.48$, $\sigmat=30\ps$,
$p_T>0.7\GeV$), and assume $\mathrm{BR}(s\to b\bar b)=1$ throughout.

\textit{Statistical procedure.}---The projected limits use a single-bin counting
likelihood
\begin{equation}
\mathcal{L}(n\,|\,\mu,\theta_b)=
\mathrm{Pois}\!\big(n\,\big|\,\mu\,s + b\,(1+\kappa_b\,\theta_b)\big)\,
\exp\!\big(-\theta_b^2/2\big),
\end{equation}
where $s$ and $b$ denote the expected signal and background yields at $3000\fbinv$, respectively. Here, $\mu$ is the signal strength---linear in $\mathrm{BR}(h\to ss)$ for the fixed assumption $\mathrm{BR}(s\to b\bar b)=1$---and $\theta_b$ is a nuisance parameter constrained by a unit Gaussian, such that $\kappa_b$ specifies the fractional background-normalization uncertainty. The expected 95\% CL upper limit is obtained from the asymptotic $\mathrm{CL}_s$ test statistic~\cite{Cowan:2010js,Read:2002hq}, evaluated on the Asimov dataset $n=b$, which also defines the $\pm1,2\sigma$ bands in Fig.~\ref{fig:limit}. For $\kappa_b=0$, the limit is purely statistical. The Monte Carlo statistical uncertainty on the optimized background tail ($\simeq7\%$, Table~\ref{tab:cutflow}) is propagated as an additional contribution to $b$.

\textit{Selection cutflow.}---Table~\ref{tab:cutflow} presents the cutflow at the benchmark point for the signal and the dominant QCD background.
\begin{table}[t]
\centering
\caption{Cutflow for a representative signal benchmark ($\ms=55\GeV$,
$\ctau=100\mm$, $2\times10^{5}$ generated events) and the dominant QCD
$b\bar bjj$ background ($2.2\times10^{6}$ generated events). The 3D selection
terminates at the $m_{\rm SV}/\Delta R$ cut; ``4D'' additionally applies the nominal timing
requirement, and ``$\dTSV>300$'' applies the optimized one. The larger QCD sample sets
the statistical precision of the optimized late-time count to $\simeq7\%$. The timing
layer rejects $93\%$ of the spatially selected QCD vertices and $24\%$ of the
signal, consistent with the quantitative content of Eq.~(\ref{eq:gain}).}
\label{tab:cutflow}
\begin{ruledtabular}
\begin{tabular}{lrr}
Selection stage & Signal & QCD $b\bar bjj$ \\
\colrule
Generated                              & 200000 & 2200000 \\
VBF tag jets                           &  53993 &  369534 \\
$\geq3$ displaced tracks               &  53884 &  322922 \\
Displaced vertex ($\geq3$ trk, $\RSV>3\mm$) & 39237 & 122288 \\
$\sum p_T>10\GeV$                      &  33583 &  101500 \\
$m_{\rm SV}/\Delta R>4$ \ (3D)         &  32145 &   96435 \\
\colrule
$+$ 4D timing (nominal)                &  24276 &    7021 \\
$+\ \dTSV>300\ps$ (opt.)               &  14916 &     200 \\
\end{tabular}
\end{ruledtabular}
\end{table}

\textit{Timed-SV efficiency.}---The four-dimensional selection requires that the displaced-vertex tracks themselves carry timing information, which is possible only within the barrel acceptance. Figure~\ref{fig:timedsv} shows, for the spatially selected (3D) vertices, the fraction that retain at least one, two, or three timed SV tracks as a function of $\ctau$; each band represents the envelope across the four scalar masses, and the line denotes their mean. The efficiency for $\geq1$ timed track is $80\text{--}87\%$ over the entire grid, with a mass spread of only a few percent and no anomalously low point. This indicates that, in this fast-simulation setup, the central SV tracks---which recoil against the forward tag jets---fall within the barrel coverage across the grid. The mild increase with $\ctau$ follows the increasing collimation and momentum of the decay products, while the slight broadening at the longest lifetime is driven by the lightest, most boosted scalar. Requiring $\geq2$ or $\geq3$ timed tracks lowers the efficiency by only a few percent, indicating that $\dTSV$ is reconstructed from several tracks rather than a single one.

\begin{figure}[h]
\centering
\includegraphics[width=0.9\linewidth]{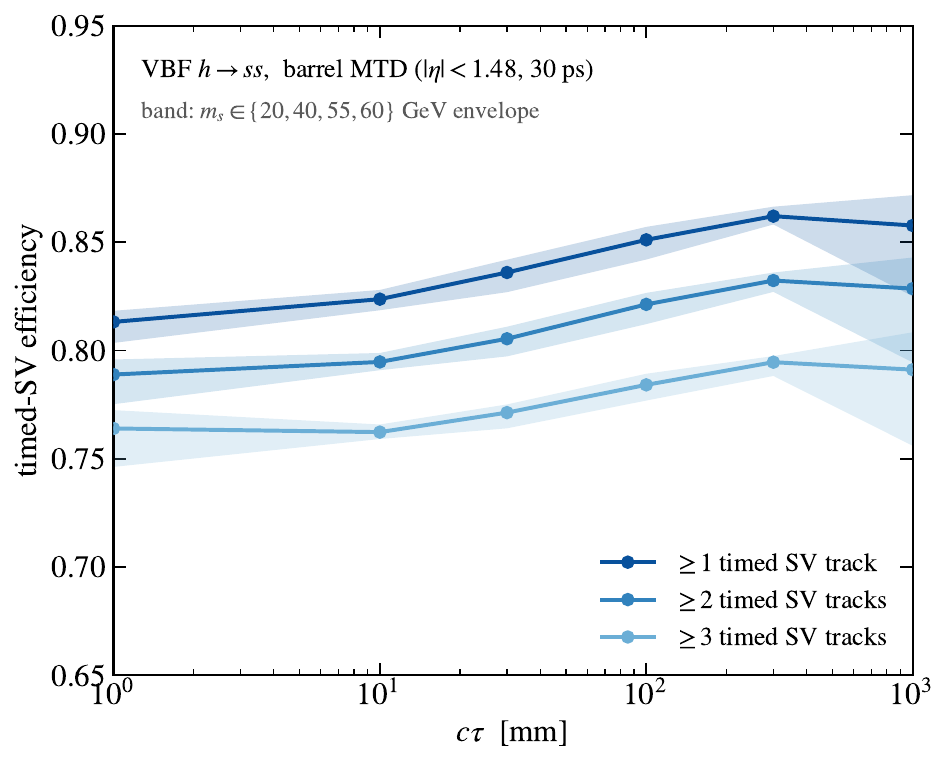}
\caption{Fraction of spatially (3D) selected displaced vertices with at least one, two, or three SV tracks timed by the barrel layer ($|\eta|<1.48$, $30\ps$), as a function of decay length. Each band represents the min--max envelope across the four scalar masses, $\ms\in\{20,40,55,60\}\GeV$, and the line denotes the mass average.}
\label{fig:timedsv}
\end{figure}

\textit{Mass-pivot decomposition.}---Because $r_B$ is common to all signal
points, the mass dependence of the gain $\Gain=r_S/\sqrt{r_B}$ resides entirely in
the signal timing survival $r_S(\ms,\ctau)$. Figure~\ref{fig:rsdecomp}(a) shows
$r_S$ at the nominal working point for the four masses. The curves intersect at
$\ctau\simeq100\mm$, where all four converge to $r_S\simeq0.74\text{--}0.76$. At
shorter lifetimes, the larger boost of the lighter scalar
($\beta\gamma\simeq p_s/\ms$) carries more decays beyond the displacement and
time thresholds, so the lighter mass has the larger $r_S$. At longer lifetimes,
the same boost drives the light scalar out of the tracker, and the heavier,
slower scalar retains the larger $r_S$. Figure~\ref{fig:rsdecomp}(b) shows the
spatial (3D) acceptance, which decreases most rapidly for the lightest scalar at long
$\ctau$, quantifying the tracker-containment leakage that drives the reversal.
The pivot in $\Gain$ is therefore the kinematic point at which
boost-enhanced timing survival and boost-induced containment loss balance,
rather than an artifact of the timing threshold.

\textit{Start-time reference availability.}---The event start time $t_0$ is
determined from prompt central reference tracks. Across the full mass--lifetime
grid, a usable reference ($N_{\rm ref}\geq1$) is found in $94\text{--}96\%$ of
signal events, with a mean multiplicity of
$\langle N_{\rm ref}\rangle\simeq15$ (Fig.~\ref{fig:nref}). The $4\text{--}6\%$
of events without a prompt reference cannot define $t_0$ and are therefore
removed only from the 4D selection, leaving the 3D baseline---and hence the
denominator of $\Gain$---unchanged. This mean multiplicity yields a statistical
start-time precision of $\sigmat/\sqrt{N_{\rm ref}}\sim8\ps$ for
$\sigmat=30\ps$.

\textit{Illustrative branching-ratio limit.}---Figure~\ref{fig:limit} shows the statistics-limited projection for $\ms=60\GeV$ discussed in the main text, for the 3D-only and optimized-4D selections; the absolute scale is an idealized lower bound set by the background normalization.
\begin{figure}[t]
\centering
\includegraphics[width=0.9\columnwidth]{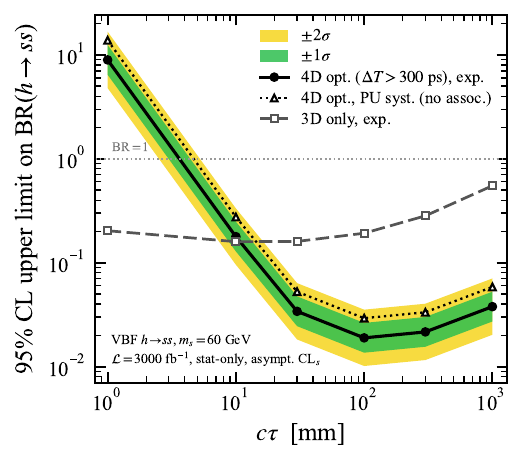}
\caption{Illustrative, statistics-limited 95\% CL upper limit on
$\mathrm{BR}(h\to ss)$, assuming $\mathrm{BR}(s\to b\bar b)=1$ (equivalently, a
limit on $\mathrm{BR}(h\to ss)\,[\mathrm{BR}(s\to b\bar b)]^2$), as a function of decay
length for $\ms=60\GeV$ at $3000\fbinv$, for the optimized 4D selection (solid,
with $\pm1,2\sigma$ bands), the same selection under the conservative
no-association pileup systematic (dotted), and the 3D-only selection (dashed).
Only statistical uncertainties are shown on the central curves; the absolute reach is an
idealized floor set by the background normalization (see text). The separation
between the solid and dotted curves brackets the pileup-association uncertainty.
The timing layer improves the limit for
$\ctau\gtrsim20\mm$ and is counterproductive for the most prompt decays.}
\label{fig:limit}
\end{figure}

\textit{Background-normalization dependence.}---Because a phenomenological
projection cannot determine the absolute QCD normalization, the limit is quoted for a
range of assumed fractional background uncertainties $\kappa_b$.
Figure~\ref{fig:limit_kappa} overlays the median upper limit at $\ms=60\GeV$ for the
3D-only and optimized-4D selections at $\kappa_b=0$ (statistics only), $1\%$, and
$5\%$. In the statistics-limited case ($\kappa_b=0$), the optimized selection
improves the limit by the gain factor ($\simeq10$ at the benchmark). Once a
normalization uncertainty is included, the spatial limit rises above unity across
the entire lifetime range. Thus, the 3D search excludes no physically allowed
branching ratio, whereas the optimized-4D limit, whose background is $\simeq500$
times smaller, remains below unity and still excludes $\mathrm{BR}\lesssim0.1$
for $\kappa_b=1\%$ ($\ctau\gtrsim10\mm$) and $\mathrm{BR}\lesssim0.7$ for
$\kappa_b=5\%$ ($\ctau\gtrsim30\mm$). At the benchmark $\ctau=100\mm$, the two
median limits differ by a factor of $\simeq216$ ($\kappa_b=1\%$) and $\simeq223$
($\kappa_b=5\%$); where both lie above $\mathrm{BR}=1$, this ratio measures
relative sensitivity rather than exclusion power. The timing advantage thus grows with
the background uncertainty---the regime expected in a real measurement---and is
what keeps the search viable. The comparison assumes the same fractional
$\kappa_b$ for the 3D and 4D selections. In a real analysis, however, $\kappa_b$ would
instead be set by the control-region statistics and transfer-factor systematics
of each selection, which need not coincide because the 4D background composition
differs from that of the 3D selection.

\begin{figure}[h]
\centering
\includegraphics[width=0.9\linewidth]{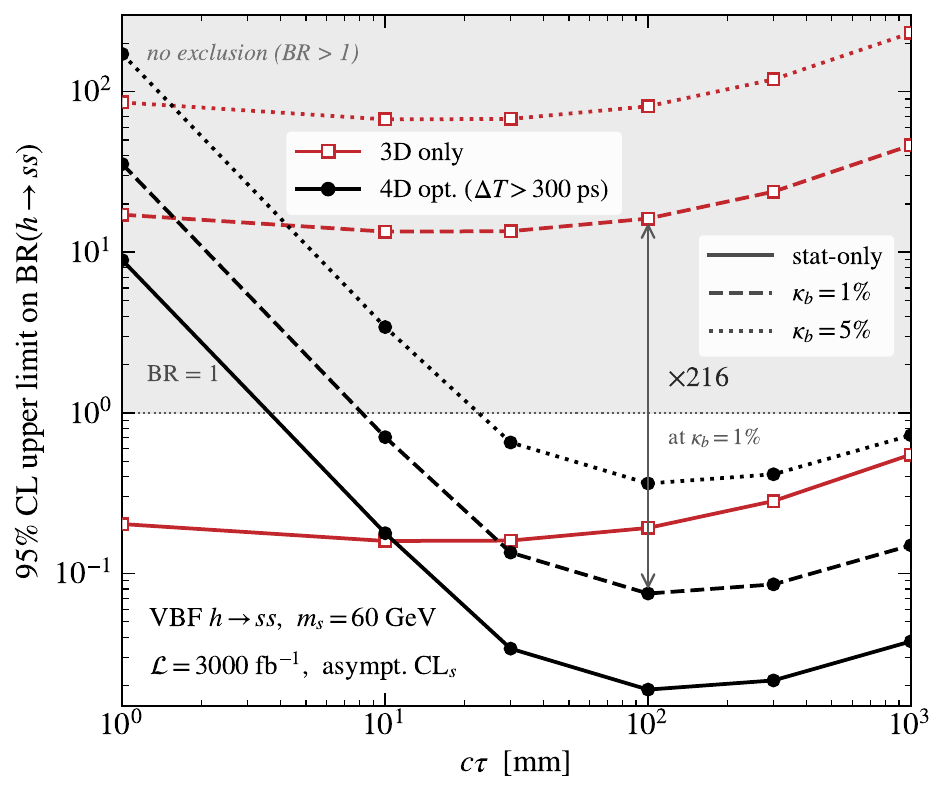}
\caption{Expected 95\% CL upper limits on $\mathrm{BR}(h\to ss)$ (assuming
$\mathrm{BR}(s\to b\bar b)=1$) as a function of decay
length for $\ms=60\GeV$ at $3000\fbinv$, for the 3D-only (red) and optimized-4D
(black) selections, each shown for an assumed background-normalization uncertainty
$\kappa_b=0$ (solid), $1\%$ (dashed), and $5\%$ (dotted). The shaded band denotes the
unphysical region $\mathrm{BR}>1$, in which an upper limit excludes no allowed
branching ratio. For $\kappa_b\geq1\%$, the entire 3D curve lies in this region,
whereas the optimized-4D curve remains below the $\mathrm{BR}=1$ boundary. The arrow indicates the
factor $\simeq216$ separating the two selections at the benchmark $\ctau=100\mm$
for $\kappa_b=1\%$. Statistical bands are omitted for clarity.}
\label{fig:limit_kappa}
\end{figure}

\textit{Velocity approximation.}---Setting $\beta_i=1$ in
$t_i^{\rm SV}=t_i^{\rm hit}-L_i/(\beta_i c)$ introduces a bias into the reconstructed production
time of a charged hadron with mass $m_i$ and momentum $p_i$ by
\begin{equation}
\Delta t_i=\frac{L_i}{c}\!\left(\frac{1}{\beta_i}-1\right)
=\frac{L_i}{c}\!\left(\frac{\sqrt{p_i^2+m_i^2}}{p_i}-1\right)
\simeq\frac{L_i}{c}\,\frac{m_i^2}{2p_i^2},
\end{equation}
a positive shift that increases for low-momentum tracks (Fig.~\ref{fig:betabias}). For the
path length to the barrel, $L\simeq1.15\,$m, the bias for a pion is $\simeq37\ps$
at $p_T=1\GeV$, $\simeq9\ps$ at $2\GeV$, and $\lesssim2\ps$ above $5\GeV$; kaons
and protons exhibit larger biases but constitute only a small fraction of the tracks from a
$b\bar b$ system. The bias is, however, \emph{common mode}: the prompt reference
tracks that determine $t_0$ and the displaced-vertex tracks are both charged hadrons
treated with $\beta=1$ and have similar momentum spectra, so it largely cancels in
$\dTSV=\langle t^{\rm SV}\rangle_{\rm SV}-t_0$, leaving only the
residual difference between the two populations---at the few-ps level, well below
$\sigmat=30\ps$. We propagate this residual by varying the correction over the
track-momentum-dependent hadron-mass envelope (pion to proton); it is included
in the systematic budget and, being much smaller than $\sigmat$, neither shifts
the break-even point appreciably nor promotes a prompt vertex beyond the
$3\sigmat$ threshold. Because the shift is positive for signal and background
vertices alike, it cannot by itself convert a prompt heavy-flavour vertex into a
coherent late signal: the common component is removed by the $t_0$ subtraction
and by the late-track-fraction requirement $\flate$, whereas a genuine long-lived
delay ($\dTSV\gtrsim100\ps$ for $\ctau\gtrsim30\mm$) far exceeds $\Delta t_i$.
Quantitatively, the same reasoning bounds the effect on the background estimate:
because the residual is only a few ps while the nominal cut is set at $3\sigmat=90\ps$,
it shifts a negligible fraction of the prompt QCD vertices---whose $\dTSV$
peaks near zero (median $25\ps$)---across the timing selection. Thus, the induced
change in the background timing-survival $r_B$ is contained within the
hadron-mass-envelope systematic and is far smaller than the pileup and
start-time systematics on $r_B$. The gain $\Gain=r_S/\sqrt{r_B}$ is therefore
unchanged at the quoted precision, and because $r_B$ is common to every mass, the
mass pivot of Fig.~\ref{fig:pivot} is left intact.

\begin{figure}[h]
\centering
\includegraphics[width=0.9\linewidth]{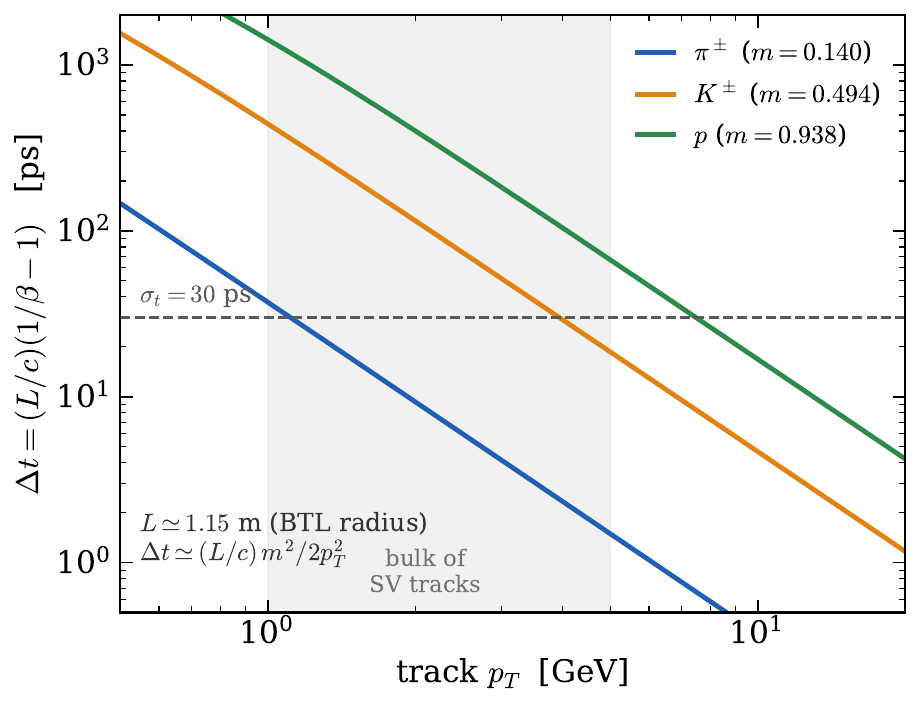}
\caption{Per-track time bias $\Delta t=(L/c)(1/\beta-1)$ arising from the $\beta=1$
approximation, shown as a function of track $p_T$, for the $\pi^\pm$, $K^\pm$,
and $p$ mass hypotheses at a barrel path length of $L\simeq1.15\,$m (for
central tracks, $p\simeq p_T$). The bias scales as $m^2/2p_T^2$ and is
largest for low-$p_T$, heavy hadrons. The dashed line indicates $\sigmat=30\ps$,
and the shaded band denotes the $p_T$ range containing the bulk of the
displaced-vertex tracks. Because the same approximation is applied to the
prompt $t_0$ reference tracks, the bias is common-mode and largely cancels in
$\dTSV$.}
\label{fig:betabias}
\end{figure}

\end{document}